\documentclass[]{iopart}

\usepackage{epsfig}

\usepackage{color}

\newcommand{\Nb}{\ensuremath{\mathrm{Nb}}}
\newcommand{\Al}{\ensuremath{\mathrm{Al}}}
\newcommand{\Si}{\ensuremath{\mathrm{Si}}}
\renewcommand{\O}{\ensuremath{\mathrm{O}}}

\renewcommand{\H}{\ensuremath{\mathrm{H}}}

\newcommand{\W}[1]{\textcolor{black}{#1}}

\begin{document}

\title{Phase qubits fabricated with trilayer junctions%
}

\author{M. Weides$^{1,2}$, R. C. Bialczak$^1$, M. Lenander$^1$, E. Lucero$^1$, Matteo Mariantoni$^1$, M. Neeley$^{1,3}$, A. D. O'Connell$^1$, D. Sank$^1$, H. Wang$^{1,4}$, J. Wenner$^1$, T. Yamamoto$^{1,5}$, Y. Yin$^1$, A. N. Cleland$^1$, and J. Martinis$^1$}
\address{$^1$Department of Physics, University of California, Santa Barbara, CA 93106, USA}
\address{$^2$Present address: National Institute of Standards and Technology, Boulder, CO 80305, USA}
\address{$^3$Present address: Lincoln Laboratory, Massachusetts Institute of Technology, Lexington, MA 02420, USA}
\address{$^4$Present address: Department of Physics, Zhejiang University, Zhejiang 310027, China}
\address{$^5$Present address: Green Innovation Research Laboratories, NEC Corporation, Tsukuba, Ibaraki 305-8501, Japan}
\ead{\mailto{martin.weides@nist.gov}, \mailto{martinis@physics.ucsb.edu}}

\date{\today}

\begin{abstract}
We have developed a novel Josephson junction geometry with minimal volume of lossy isolation dielectric, being suitable for higher quality trilayer junctions implemented in qubits. The junctions are based on in-situ deposited trilayers with thermal  tunnel oxide, have micron-sized areas and a low subgap current. In qubit spectroscopy only a few avoided level crossings are observed, and the measured relaxation time of $T_1\approx400\;\rm{nsec}$ is in good agreement with the usual phase qubit decay time, indicating low loss due to the additional isolation dielectric.
\end{abstract}


\maketitle

\section{Introduction}

The energy relaxation time $T_1$ of superconducting qubits is affected by dielectric loss,  nonequilibrium quasiparticles \cite{MartinisPRL09}, and charge or bias noise, and varies between a few nano- to several microseconds, depending on qubit type, material, and device layout. Superconducting qubits are commonly based on $\Al$ thin films, and their central element, the non-linear inductor given by a Josephson tunnel junction (JJ), is formed either by overlap \cite{SteffenPRL06} or window-type geometries \cite{KlineSST09}. Qubit spectroscopy reveals coupling to stochastically distributed two-level systems (TLSs) in the tunnel oxide \cite{SimmondsPRL04,LupascuPRB09,LisenfeldPRB10,Bushev_PRB10,DeppePRB07}
which provide a channel for qubit decoherence. While the physical nature of TLSs is still under debate, their number was shown to decrease with junction size and their density with higher atomic coordination number of the tunnel oxide \cite{KlineSST09,MartinisPRL05}. The number of coherent oscillations in the qubit is limited by, among other decoherence mechanisms such as nonequilibrium quasiparticles, the \emph{effective} dielectric loss tangent $\tan\delta_{\rm{eff}}$ \cite{MartinisPRL05}. The overlap geometry provides JJs with amorphous barriers with no need for isolation dielectrics, being itself a source for additional TLSs and dielectric losses. The window geometry is used for higher quality, e.g. epitaxial, trilayer JJs with in-situ grown barriers. Besides complex fabrication, they have the drawback of requiring additional isolation dielectrics \cite{BaronePaterno}.
The importance of keeping the total dielectric volume in qubits small to reduce the additional loss was shown in Ref. \cite{MartinisPRL05}.\par
In this paper we give an overview of our standard technology for junction fabrication, and present an alternative junction based on sputtered trilayer stacks, which provide an intrinsically cleaner tunnel oxide and is well suited for micron-sized trilayer qubit junctions. The so-called \emph{side-wall passivated JJs} provide contact to the top electrode without adding too much lossy dielectric to the circuitry, which would negatively affect the loss tangent. The trilayer isolation is achieved via an electrolytic process. These novel JJs were realized in a flux-biased phase qubit and characterized by i) current transport measurements on reference junctions and ii) spectroscopy and time-domain measurements of the qubit.\par

By systematically replacing only the Josephson junction, being central to any superconducting qubit, we aim to analyze the loss contributions of this specific element, and, ideally, develop low-loss Josephson junctions for superconducting qubits and improve our qubit performance. We found performance comparable to the current generation of overlap phase qubits.

\section{Novel geometry}

\begin{figure}[tb]
\begin{center}
\includegraphics[width=8.6cm]{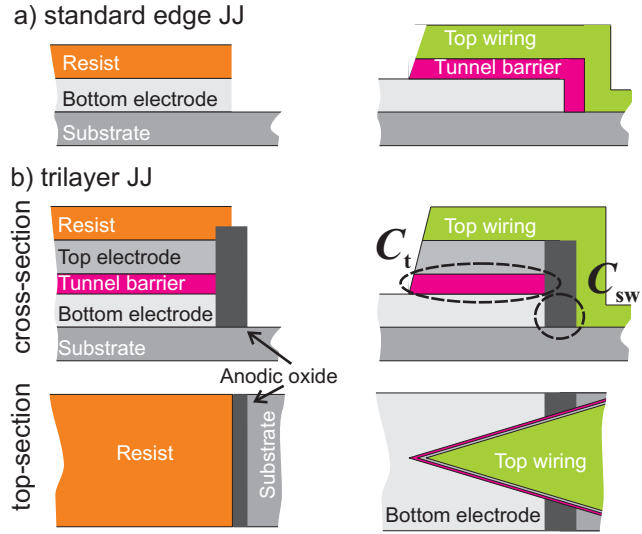} \caption{(Color online) Schematics of the a) overlap JJ and b) side-wall passivated JJ, offering minimal volume of passivation region. Left (right) part: before (after) the top-layer deposition. After the edge etch in the trilayer stack, the side-wall oxide is grown by anodic oxidation. The trilayer JJ has in-situ grown tunnel oxides to avoid sources of residual impurities. Patterning of the top wiring and etching below the tunnel barrier yields the tunnel junction.}\label{SidewallAnod}
\end{center}
\end{figure}

Figure \ref{SidewallAnod} depicts the patterning process for our standard overlap (a) and trilayer junctions (b). Our standard process has an oxide layer grown on an ion mill cleaned aluminum edge, which was previously chlorine etched. The top wiring is then etched back below the oxide layer using argon with $\sim 10\%$ chlorine mixture. For the trilayer process, the in-situ sputtered $\Al$-$\Al\O_x$-$\Al$ trilayer has a thermally grown tunnel oxide barrier, formed for 10 min at $140\:\rm{mTorr}$ at room temperature. After deposition of the trilayer stack an edge is etched. The bottom electrode of the trilayer stack is isolated from the top electrode wiring by a self-aligned nanometer thin dielectric layer, grown for $\Al$ (or other suitable electrode metals such as $\Nb$) by anodic oxidation \cite{Kroger81SNAP}. The metallic aluminium serves as partly submerged anode in a liquid electrolytic mixture of $156\;\rm{g}$ ammonium pentaborate, $1120\;\rm{ml}$ ethylene glycol
and $760\;\rm{ml}$ $\H_2\O$ at room temperature. A gold-covered metal served as cathode and the electric contact was made outside the electrolyte to the anode. By protecting parts of the aluminum electrode with photoresist only a well-defined area was oxidized by passing a constant current through the Al film and converting the metallic surface to its oxide form. The oxide thickness can be controlled by the voltage drop across the electrolyte. After a light ion clean and top wiring deposition the resist is patterned to define the junction area. Finally, the trilayer is etched below the tunnel barrier, yielding Josephson junctions with planar tunnel barrier and isolation dielectric on just one side of the tunnel area. \W{For $\Nb$ junctions a similar patterning process, without minimizing the dielectric loss contribution, was developed using anodic $\Nb$ oxide and covered by $\Si\O_2$ \cite{Mueller_01}.} The in-situ grown tunnel oxide avoids sources of residual impurities such as hydrogen, hydroxide or carbon at the interface vicinity, which may remain even after ion-milling in our standard process. These trilayer junctions are fully compatible with our standard process using overlap patterning and no junction side-wall.\\

\subsection{Transport}
\begin{figure}[tb]
\begin{center}
\includegraphics[width=8.6cm]{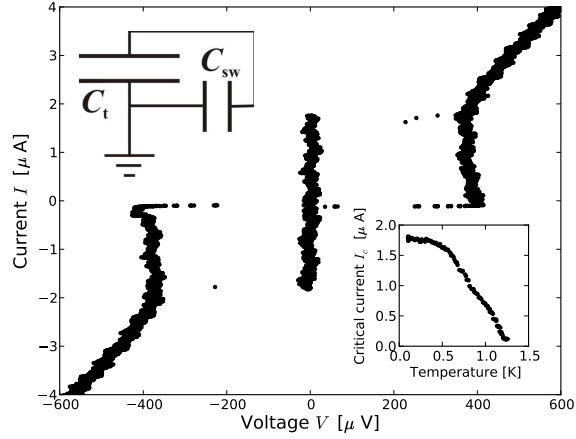} \caption{Current-voltage-characteristic at $100\;\rm{mK}$ and $I_c(T)$ dependence (lower inset) of a $3\;\rm{\mu m^2}$ side-wall passivated trilayer junction. Top inset: dielectric circuit elements of the junction. The tunnel oxide capacitance $C_{\rm{t}}$ is connected in parallel with the capacitor formed by the side-wall oxide $C_{\rm{sw}}$.} \label{IVC}
\end{center}
\end{figure}

Transport measurements on a $\sim 3\:\rm{\mu m^2}$ reference junction at $100\;\rm{mK}$ are shown in Fig. \ref{IVC}. The critical current $I_c$ is $1.80\:\rm{\mu A}$, with normal resistance $R_n=150\rm{\Omega}$ yielding $I_c R_n=270\rm{\mu V}$, close to the calculated Ambegaokar-Baratoff value of $I_c R_n=298\rm{\mu V}$ for the measured superconducting gap of $190\;\rm{\mu V}$. \W{The back bending of the voltage close to the gap voltage is attributed to self-heating inside the junction.} The retrapping current of $\approx 0.01 \cdot I_c$ indicates a very small subgap current. The current transport is consistent with tunneling, and we can exclude transport via metallic pinholes, located in the $\sim5\;\rm{nm}$ thin side-wall dielectric. As a further check, the $I_c(T)$ dependence is as expected, see inset in Fig. \ref{IVC}, with a critical temperature $T_c$ of $1.2\;\rm{K}$.\\

\section{Measurement}

The qubit is a flux-biased phase qubit that is coupled via a tunable mutual inductance to the readout-SQUID \cite{NeeleyPRB08}. \W{The total qubit capacitance $C_{total}$, see upper inset of Fig. \ref{Fig:2dSpectro}, is given by the tunnel oxide $C_{\rm{t}}$, the anodic side-wall oxide $C_{\rm{sw}}$ and shunt capacitor $C_{\rm{s}} \approx 1250\;\rm{fF}$ dielectric}, provided by a parallel plate capacitor with relative
permittivity $\epsilon' \simeq 11.8$ made from hydrogenated amorphous silicon (a-Si:H). The measurement process follows the standard phase qubit characterization \cite{MartinisSC_PhaseQubits09}.

\subsection{Spectroscopy}

When operated as a qubit, spectroscopy over a range of more than $2.5\;\rm{GHz}$ revealed clean qubit resonance spectra with just two avoided level crossings of $40$--$50\;\rm{MHz}$ coupling strength (at $6.96$ and $7.32\:\rm{GHz}$, as shown in Fig. \ref{Fig:2dSpectro}). The excitation pulse length is $1\;\rm{\mu sec}$, and the qubit linewidth is about $3\:\rm{MHz}$ in the weak power limit. The qubit visibility, measured in a separate experiment, is about $86\%$, which is in the range we found for our standard phase qubits.\\
Qualitatively, the TLS number and coupling strength per qubit is lower than in other trilayer systems \cite{KlineSST09}, that have larger tunnel areas. The TLS density per qubit has roughly the same order of magnitude as in conventional overlap qubits with similar tunnel area dimensions \cite{SteffenPRL06}.

\subsection{Relaxation}

\begin{figure}[tb]
\begin{center}
\includegraphics*[width=8.6cm]{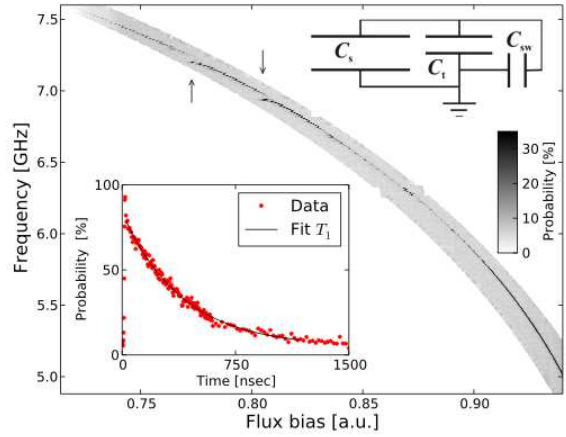} \caption{(Color online) 2D spectroscopy of side-wall passivated trilayer qubit at $25\;\rm{mK}$. Two avoided level crossings due to qubit-TLS coupling are observed at $6.96$ and $7.32\:\rm{GHz}$ (arrows). Top inset: Dielectric circuit schematics of the qubit. Bottom inset: Qubit relaxation measurement.}\label{Fig:2dSpectro}
\end{center}
\end{figure}

Qubit relaxation measurements via $\pi$ pulse excitation and time-varied delay before readout pulse were obtained when operated outside the avoided level structures. We measured a relaxation time $T_1$ of about $400\;\rm{nsec}$, as shown in the lower inset of Fig. \ref{Fig:2dSpectro}.
This relaxation time is similar to that observed in the overlap qubits, which consistently have $300\textrm{--}500\:\rm{nsec}$ for $\sim2$-$4\;\rm{\mu m^2}$ JJ size. Apart from the change to trilayer junctions, no modification from the previous design was made.

\section{Loss estimation}
\begin{table}[tb]
\begin{tabular}{lllllc}
  \hline
  \hline
  dielectric elements & &capacitance  & loss $\tan\delta_i$&$\frac{C_i}{C_{total}}\tan\delta_i$&\\
  & &$[\rm{fF}$] & &&\\
  \hline
 shunt capacitor a-Si:H&$C_{\rm{s}}$  & 1250& $2\cdot10^{-5}$&$1.83\cdot10^{-5}$& \cite{oConnellAPL}\\
  anodic side-wall oxide& $C_{\rm{sw}}$ & 3.2 & $<1.6\cdot10^{-3}$&$<3.7\cdot10^{-6}$& \cite{MartinisPRL05}\\
 tunnel barrier& $C_{\rm{t}}$  & 116&$<1.6\cdot10^{-3}$&$<1.36\cdot10^{-4}$&\cite{MartinisPRL05}\\
 \emph{measured} $\tan\delta_{m}$ &&&&$6.6\cdot 10^{-5}$&\\
  \hline
  \hline
\end{tabular}\caption{Dielectric parameters for anodic oxide, shunt capacitance, and tunnel barrier. $\tan\delta$ is given for low temperature and low power at microwave frequencies. The capacitance for the tunnel oxide $\Al\O_x$ is taken and corrected for $\Al$ electrodes from Ref. \cite{ZantAPL94} for the dimensions given in the text. For qubits the loss tangent is calculated away from TLS resonances, as the losses in small size anodic side-wall oxide and tunnel barrier are smaller than the bulk value considered for the specific loss contribution $\frac{C_i}{C_{total}}\tan\delta_i$. The \W{measured loss} $\tan \delta_{m}$ is a factor 2-3 smaller than $\tan \delta_{\rm{eff}}$, the weighted sum of all specific loss contributions. \label{Tab:LossEstimation}}
\end{table}

We estimate the additional dielectric loss due to the sidewall oxide.
The \emph{effective} loss tangent of a parallel combination of capacitors is given by
\[
\tan{\delta_{\rm{eff}}}=
\frac{\epsilon''_{\rm{eff}}}{\epsilon'_{\rm{eff}}}=
\frac{\sum\limits_{i} \epsilon''_i \frac{A_i}{d_i}}
{\sum\limits_{i} \epsilon'_i \frac{A_i}{d_i}}=
\frac{\sum\limits_{i}C_i \tan\delta_i}{\sum\limits_{i} C_i}
\]
with $\epsilon'_i$ and $\epsilon''_i$ being the real and imaginary part of the individual permittivity for capacitor $i$ with area $A_i$ and dielectric thickness $d_i$.

Now, we discuss the individual loss contributions for all dielectrics. We design the qubit so that the dominant capacitance comes from the shunt capacitor made from a-Si:H, which has a relatively low loss tangent of $2 \cdot 10^{-5}$.  Including the non-negligible capacitance of the tunnel junction, this gives an effective loss tangent to the qubit of $1.83 \cdot 10^{-5}$. Because the junction capacitance is about 10\% of the shunting capacitance, the effective junction loss tangent is 10 times less than the loss tangent of the junction oxide.  We statistically avoid the effects of two-level systems by purposely choosing to bias the devices away from the deleterious resonances.  The loss tangent of the junction is smaller than the value for bulk aluminum oxide, approximately $1.6 \cdot 10^{-3}$, and probably smaller than $5\cdot 10^{-5}$ since long energy decay times ($500\;\rm{nsec}$) have been observed for an unshunted junction when operated away from resonances \cite{MartinisPRL05}.\\
The anodic side-wall oxide contributes a small capacitance of about 3.2\,fF, which can be calculated assuming a parallel plate geometry.  Here, we use the dielectric constant $\epsilon' = 9$ for aluminum oxide, assume an area given by $2\,\mu$m, the width of the overlap, multiplied by $0.1\mu$m the thickness of the base layer, and estimate the thickness of the oxide  $\simeq5\,$nm as determined by the anodic process \cite{Kroger81SNAP}. The anodic oxide is assumed to have a bulk loss tangent of $1.6 \cdot 10^{-3}$ \cite{oConnellAPL}, which gives a net qubit loss contribution of $3.7 \cdot 10^{-6}$, about 5 times lower than for the a-Si:H capacitor.  Note that we expect the loss from this capacitance to be even lower because of statistical avoidance of the TLS loss \cite{MartinisPRL05}.  The small volume of the capacitor, equivalent to a $\sim 0.5\,\mu\textrm{m}^2$ volume tunnel junction, implies that most biases do not put the qubit on resonance with two-level systems in the anodic oxide.

\section{Qubit lifetime and effective loss tangent}

From the measured energy decay time $T_1=400\;\rm{nsec}$ , we determine the loss tangent of the qubit to be $\delta_m = (T_1\;\omega_{10})^{-1} \approx 6.6\cdot 10^{-5}$, using a qubit frequency $\omega_{10}/2\pi=6\;\rm{GHz}$.  This is 3-5 times larger than our estimation of our dielectric losses, as shown in Table \ref{Tab:LossEstimation}.  We believe the \W{qubit} dissipation mechanism comes from some other energy loss sources as well, such as non-equilibrium quasiparticles  \cite{MartinisPRL09}.\\

\section{Conclusion}
In conclusion, we have shown that the use of a anodic oxide, self-aligned to the junction edge, does not degrade the coherence of present phase qubits \cite{MartinisSC_PhaseQubits09}. We found performance comparable to the current generation of overlap phase qubits.

The new junction geometry may provide a method to integrate submicron sized, superior quality junctions (lower TLS densities) grown, for example, by MBE epitaxy to eliminate the need for shunt dielectrics. Also, our nanometer-thin, three dimensional-conform anodic passivation layer can be replaced by a self-aligned isolation dielectric at the side-wall, which could be used for all types of trilayer stacks.

Devices were made at the UCSB Nanofabrication Facility, a part of the NSF-funded National Nanotechnology Infrastructure Network.

The authors would like to thank D. Pappas for stimulating discussions. This work was supported by IARPA under grant W911NF-04-1-0204. M.W. acknowledges support from AvH foundation and M.M. from an Elings Postdoctoral Fellowship.

\end{document}